\documentclass[aps,pre,nopacs,superscriptaddress,groupedaddress]{revtex4}
\usepackage{graphicx}  
\usepackage{dcolumn}   
\usepackage{amssymb}   
\usepackage{amsmath}
\usepackage{color}

\begin{document}



\title{Emergence of synchronised and amplified oscillations in neuromorphic networks with long-range interactions}
\author{I. Apicella}
\affiliation{Dipartimento di Fisica `G. Galilei', INFN, Universit\'a di Padova, Via Marzolo 8, 35131 Padova, Italy}
\affiliation{ Dipartimento di Fisica  `E.R. Caianiello',  INFN gruppo coll. di Salerno, Universit\'a di Salerno, Via Giovanni Paolo II, I84084, Fisciano (SA),
Salerno, Italy}
\author{D. M. Busiello}
\affiliation{Ecole Polytechnique F\'ed\'erale de Lausanne (EPFL), Institute of Physics Laboratory of Statistical Biophysics, 1015 Lausanne, Switzerland}
\author{S. Scarpetta}
\affiliation{ Dipartimento di Fisica  `E.R. Caianiello',  INFN gruppo coll. di Salerno, Universit\'a di Salerno, Via Giovanni Paolo II, I84084, Fisciano (SA),
Salerno, Italy}
\author{S. Suweis}
\affiliation{Dipartimento di Fisica `G. Galilei', INFN, Universit\'a di Padova, Via Marzolo 8, 35131 Padova, Italy}
\affiliation{Padova Neuroscience Center, Universit\'a di Padova, Via Orus, 2, 35131 Padova, Italy}
\begin{abstract}

Neuromorphic networks can be described in terms of coarse-grained variables, where emergent sustained behaviours spontaneously arise if stochasticity is properly taken in account. For example it has been recently found that a directed linear chain of connected patch of neurons amplifies an input signal, also tuning its characteristic frequency. 
Here we study a generalization of such a simple model, introducing heterogeneity and variability in the parameter space and long-range interactions, breaking, in turn, the preferential direction of information transmission of a directed chain. On one hand, enlarging the region of parameters leads to a more complex state space that we analytically characterise; moreover, we explicitly link the strength distribution of the non-local interactions with the frequency distribution of the network oscillations. On the other hand, we found that adding long-range interactions can cause the onset of novel phenomena, as coherent and synchronous oscillations among all the interacting units, which can also coexist with the amplification of the signal. 

\end{abstract}

\maketitle


\section{Introduction}

The brain is a self-organising and evolving network of fundamental processing elements (the neurons) and it is configured differently from classical von Neumann' architectures. At the microscopic level, the neurons are the elementary computational units forming complex networks wired through axons, dendrites and synapses, i.e. specialised contacts transmitting information. Brain tasks, such as encoding and decoding information from the external world and elaborating behavioural responses, are developed from the collective activity of neuronal populations \cite{mascaro1999effective,sanger2003neural,chapin2004using}. 

A remarkable signature of these coordinated dynamics are neuronal avalanches \cite{beggs2004neuronal}. The cerebral cortex is never silent, not even under resting conditions nor in the absence of stimuli. In fact, it exhibits a state of spontaneous heterogeneous and yet correlated activity \cite{mantini2007electrophysiological}. In a seminal work, J. Beggs and D. Plenz \cite{beggs2003neuronal} succeed at  resolving the internal spatiotemporal organisation of the outbursts of neuronal activity by analysing neuronal cultures as well as acute slices of rat cortex and recorded spontaneous local field potentials (LFPs). These bursts of activity then resulted in cascade of successive local events, organised as neuronal avalanches, interspersed by periods of quiescence. These avalanches have been observed in many other subsequent studies \cite{pasquale2008self,petermann2009spontaneous,lombardi2012balance}, and this emergent property is similar to that one seen in many other complex systems.  In fact, just to cite some examples, events like earthquakes, forest fires, and nuclear chain reactions emerge as one unit activates and causes other units to do so in turn, thereby initiating a cascade that propagates through the larger system \cite{bak2013nature}. 

Another widespread neuronal collective phenomena observed in brains of different species are sustained oscillations of neural activity, e.g., rhythmic patterns of spiking neurons  in the central nervous system. These oscillations can be observed and measured, e.g. through electroencephalogram (EEG) \cite{herrmann2016eeg,massobrio2015criticality} and synchronous brain activity is also observed \cite{buzsaki2006rhythms}. 
This oscillatory activity of large groups of neurons may be explained in various ways \cite{mejias2011emergence,lombardi2014temporal}, but it is known that excitatory-inhibitory interactions are crucial in order to induce oscillations of their firing patterns \cite{dayan2003theoretical}. Moreover, depending on the properties of the connections, such as the coupling strengths, time delay, etc... the spike trains of the interacting neurons may become synchronised \cite{borgers2003synchronization}.
 
The immense complexity of brain network and dynamics invites thinking in terms of simple physical modelling approaches, that can however display very complex behaviour and thus enlighten underlying main driving processes \cite{chialvo2010emergent,tagliazucchi2012criticality,rocha2018homeostatic,odor2019critical,scarpetta2018hysteresis}. For example, many models have investigated the role of the network structures in promoting oscillatory synchronised activity in neural networks \cite{motter2005network,arenas2008synchronization}. In oscillatory networks, where neurons are modelled by simple oscillators, the ability to synchronise is generally enhanced in small-world networks as compared to regular lattices \cite{barahona2002synchronization}. Therefore, the topological properties of the interaction network play an important role in driving the neural activity, especially when the neural dynamics is described in terms of simple and essential features.

The Wilson and Cowan model is one of the well-established simplified models of neuronal network dynamics at the mesoscopic level \cite{wilson1972excitatory}. It incorporates individual elements as two-state stochastic oscillators with one quiescent state and one excited state, and random transitions between these two states which are influenced by the mutual coupling among the network elements. The model has been introduced in the deterministic limit of many coupled elements and neglecting the role of the intrinsic noise (i.e. mean field approach). In fact, in spite of its simplicity, the Wilson-Cowan model may describe several dynamical behaviours as a function of its parametrization, qualitatively reproducing experimentally observed regimes of neural dynamics \cite{destexhe2009wilson}.

Recently, the previously neglected mesoscopic noise effects have been incorporated in this model to take in account the effect of the finite size of networks and circuits \cite{cowan2016wilson}. Considering such stochasticity in the Wilson-Cowan is very important to describe both neuronal avalanches \cite{cowan2016wilson}, up and down states \cite{hidalgo2012stochastic} and oscillations \cite{fanelli2017noise}.  Stochastic amplification \cite{mckane2005predator} have been first proposed as a viable approach to the study of spontaneously generated, regular and sustained oscillations in natural systems by noise, e.g., in brain activity \cite{hidalgo2012stochastic}. However such oscillations produced by this effect are very small in amplitude. This limitation was recently overcome by considering an asymmetric directed couplings through a linear chain neuromorphic network driving a further amplification mechanism that eventually results in large oscillations (exponentially amplified) across the chain \cite{fanelli2017noise}. However, the proposed model did not investigate the role of more biological parametrisation such as difference in density between excitatory and inhibitory cells, as well as heterogeneity in the coupling strengths. Moreover in order to achieve synchronisation, a fine tuning of the different model parameters must be performed. 

In this work  we study a generalization of such stochastic Wilson-Cowan model, investigating a neuromorphic network constituted by $\Omega$ nodes of inhibitory and excitatory neuronal populations, arranged on a linear chain backbone, but also connected through long-range backward interactions, breaking, in turn, the preferential direction of information transmission of the directed chain. On one hand, enlarging the region of parameters leads to richer states phase space characterising the system's dynamics behaviours. Moreover, we explicitly link the strength distribution of the non-local interactions with the frequency distribution of the network oscillations. On the other hand, adding long-range interactions can cause the onset of novel phenomena, as coherent and synchronous oscillations among all the interacting units, and coexisting with the amplification of the signal. 

The paper is organized as follow:  in the next section we present the analytical framework describing the stochastic neuromorphic network dynamics. In section \ref{sectionResults}A  we consider a directed linear chain and we analytically characterize the effects of the  heterogeneity in the model parameters; Subsection \ref{sectionResults}B presents the main result of this work: adding directed long-range links to the linear backbone of the neuromorphic network leads to synchronisation and phase coherence, novel phenomena not observed in the directed chain. Finally, we show how a local condition of the network connectivity may have a strong impact on the amplification of the oscillatory signal.

\section{The Model}
The Wilson-Cowan (WC) model provides a standard framework to describe the interactions between mesoscopic competing units. Its most prominent application concerns a coarse-grained description of neuronal networks, in which excitatory and inhibitory units are described in terms of their concentrations in a given region \cite{wilson1972excitatory,cowan2016wilson}. In the same spirit of  \cite{fanelli2017noise}, we consider a network of a finite number $\Omega$ of nodes, each of them constituting a patch of volume $V_i$. Within a given patch $i$, an excitatory and an inhibitory units can interact, being characterised by a number of artificial neurons $X_i$ and $Y_i$, respectively. Their interactions can be expressed in terms of a birth (activation) and death (de-activation) stochastic process: with probability $P(1\rightarrow0)=1$ a neuron stops to fire, while with probability $P(0\rightarrow1)=F(s)$ a new neuron spikes. The dynamics of a single unit $i$ can be thus described through the following stochastic transitions:

\begin{gather}
X_i \xrightarrow{1} \emptyset \qquad \emptyset \xrightarrow{F[s_{X_i}]} X_i \nonumber \\
Y_i \xrightarrow{1} \emptyset \qquad \emptyset \xrightarrow{F[s_{Y_i}]} Y_i \nonumber
\end{gather} 
where $F[s_{X}] = g[X] f[s_X]$ is usually called ``activation function", with $f[s] = 1/(1+e^{-s})$ a sigmoidal function encoding the saturation behaviour to external stimuli \cite{fanelli2017noise}.

\begin{figure}
\begin{center}
\includegraphics[scale=0.2]{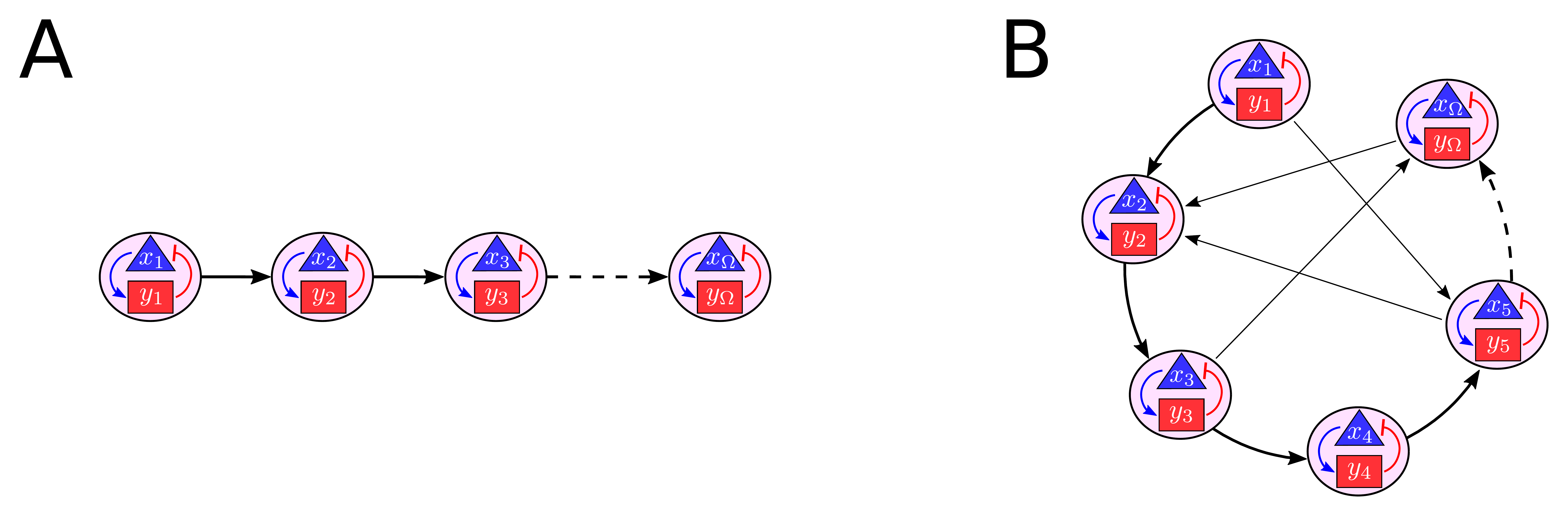}
\caption{Schematic representation of a neuromorphic network with $\Omega$ nodes, constituted by excitatory (blue triangles) and inhibitory (red squares) populations, like in Wilson-Cowan model for neuronal networks. The simplest topology is the directed linear chain (Panel A).
 In Panel B a more  complex structure, with the addition of long-range links to the directed linear chain, is shown. \label{modello}}
\end{center}
\end{figure}

The arguments of $F[s]$ are defined as follows \cite{fanelli2017noise}:
\begin{eqnarray}
s_{x_i}&=& -r_i(y_i - p_i) + D_i \sum_j L_{ij}x_j - D_i \sum_j L_{ij}y_j \nonumber \\
s_{y_i}&=& r_i(x_i - (1-p_i)) + D_i \sum_j L_{ij}x_j - D_i \sum_j L_{ij}y_j
\label{s}
\end{eqnarray}
with $x_i= \frac{n_{X_i}}{V_i}$  and $y_i= \frac{n_{Y_i}}{V_i}$ the concentration of units (excitatory or inhibitory) on each node ($n_{X_i}$ and $n_{Y_i}$ are the number of active artificial neurons of type $X_i$ or $Y_i$ respectively). As defined in (\ref{s}), $s_{x_i}$ and $s_{y_i}$ contain information about the local interactions, through the control parameter $r_i$ and about the non-local interactions, through the positive coupling constant $D_i$. The Laplacian matrix $L_{ij}=A_{ij}-\sum_k A_{ki}\delta_{ij}$ is the discrete analog of the diffusion operator in the continuous space through the network described by the adjacency matrix $A_{ij}$.

In each region $i$, the above equations model the spiking neuron dynamics and the related level of the concentrations of excitatory and inhibitory units. The dependence on $i$ in each parameters ($r_i$, $p_i$ and $D_i$) mimics the fact that the connections between and within each region are heterogeneous. The matrix $A$ represents the interactions between the different regions and the Laplacian matrix describes how information diffuses on the network, through both short and long-range interactions. Despite its streamlined dynamics, this generalized neuromorphic model has a very rich and complex dynamical behaviour.

The microscopic stochastic dynamics can be translated into a corresponding Master Equation \cite{gardiner2009stochastic} whose solution gives the probability that $n^{i}_x$ excitatory and $n^{i}_y$ inhibitory neurons are spiking at time $t$ in regions $i=1,...,\Omega$. This solution cannot be obtained analytically and in order to describe the system at a coarse-grained level, i.e. in terms of the concentrations, we perform the Kramers-Moyal expansion, leading to the following set of coupled Langevin equations  \cite{gardiner2009stochastic}:

\begin{eqnarray}
\frac{d}{dt} x_i&=& \frac{1}{\gamma_i} (F[s_{x_i}]-x_i) + \frac{1}{\gamma_i \sqrt{V1}} \sqrt{F[s_{x_i}]+x_i} \lambda_i^{(1)} \nonumber \\
\frac{d}{dt}  y_i&=& \frac{1}{\gamma_i} (F[s_{y_i}]-y_i) + \frac{1}{\gamma_i \sqrt{V1}} \sqrt{F[s_{y_i}]+y_i} \lambda_i^{(2)}
\label{langevin}
\end{eqnarray}
where $V_i$ is the volume of the $i$-th patch and $\gamma_i = V_i/V_1$. Moreover, the noise contributions are encoded in $\lambda_i^{(1)}$ and $\lambda_i^{(2)}$, which are Gaussian distributed with zero mean and correlator $<\lambda_i^{(l)}\lambda_j^{(m)}>=\delta_{ij}\delta_{lm}\delta(t-t')$. \\
Notice that setting
\begin{equation}
F[X_i] = (1-p_i) (1 + e^{-2D_i (p_i-p_{i-1})})f[s_{X_i}] \qquad F[Y_i] = p_i (1 + e^{-2D_i (p_i-p_{i-1})})f[s_{Y_i}]
\end{equation}
the parameter $p_i$ (respectively $1-p_i$) can be interpreted as the fraction of inhibitory (excitatory) units in the node $i$ at equilibrium. In other words, $x^{*}=1-p_i$ and $y^{*}=p_i$ are - by construction - the fixed points of the deterministic dynamics, obtained performing the thermodynamic limit $V_i \to + \infty$ for all $i=1,...,\Omega$.

The simplest topology that we can consider in this framework is a directed linear chain (DLC), sketched in Figure \ref{modello}A, with $r_i$, $D_i$ and $p_i$ constant for all the nodes (i.e., neglecting the dependence on $i$). Moreover, as a further simplification it is possible to set  $p_i=1/2$. This latter case has been extensively analysed in \cite{fanelli2017noise}, finding that sustained oscillation can appear due to stochastic resonance \cite{mckane2005predator}, and eventually they get amplified through the chain. Moreover, changing the ratio between volumes of the patches, the frequency of the signal in each node can be externally controlled \cite{fanelli2017noise}.

In order to add complexity to this picture, we proceed exploring two different directions: $a)$ we consider a DLC with values of the local and non-local parameters $r_i>0$ and $D_i>0$ different for each node $i$, while keeping $p_i = p \neq 1/2$, $\forall i$; $b)$ we increase the topological complexity adding long-range interactions to the directed linear chain, breaking, in turn, the preferential direction of information transmission of a directed chain (see Figure \ref{modello}B for an illustrative example).

As presented below, in the first case more complex conditions for the onset of amplified oscillations are obtained. Moreover, the variability of the diffusion coefficients among the nodes allows for an arbitrary tuning of the distribution of the response frequencies. In the second case, the amplification is not the only emergent phenomena that we observe. In fact, we find a region of the parameter space where neurons activity starts synchronising: the vast majority of them oscillate with the same  dominant frequency and, often, with robust phase locking. 
 
\section{Results}
\label{sectionResults}
\subsection{Stability of an heterogeneous linear chain.}

In this section we want to focus on the consequences of having heterogeneous values for the interaction strengths $r_i$ and $D_i$, governing the interactions within and between units, respectively. To this aim, we specialize the analysis to a directed linear chain, keeping $p_i=p$ $\forall i$, i.e. the proportion of inhibitory and excitatory artificial  neurons in each unit is the same.

The effects of the stochasticity is investigated under the Linear Noise Approximation (LNA) \cite{van1992stochastic}. The noise acts as a first order perturbation to the deterministic fixed points, that is $x_i=x^{*}+\frac{\xi_i}{\sqrt{V_i}}$ and $y_i=y^{*}+\frac{\eta_i}{\sqrt{V_i}}$ in the Langevin equations (\ref{langevin}). Performing the expansion up to the first order in $\frac{1}{\sqrt{V_1}}$, we end up with the following Langevin equations for the fluctuations $\vec{\zeta} = ( \vec{\xi}, \vec{\eta} )$:

\begin{equation}
\frac{d\vec{\zeta}}{dt}=\textbf{J}\vec{\zeta}+\textbf{B}\vec{\lambda}
\label{zeta}
\end{equation}

where \textbf{J} is the Jacobian of the dynamics and \textbf{B} the matrix containing the noise terms (further details can be found in Appendix A).

\begin{figure}[t]
\includegraphics[width=15 cm]{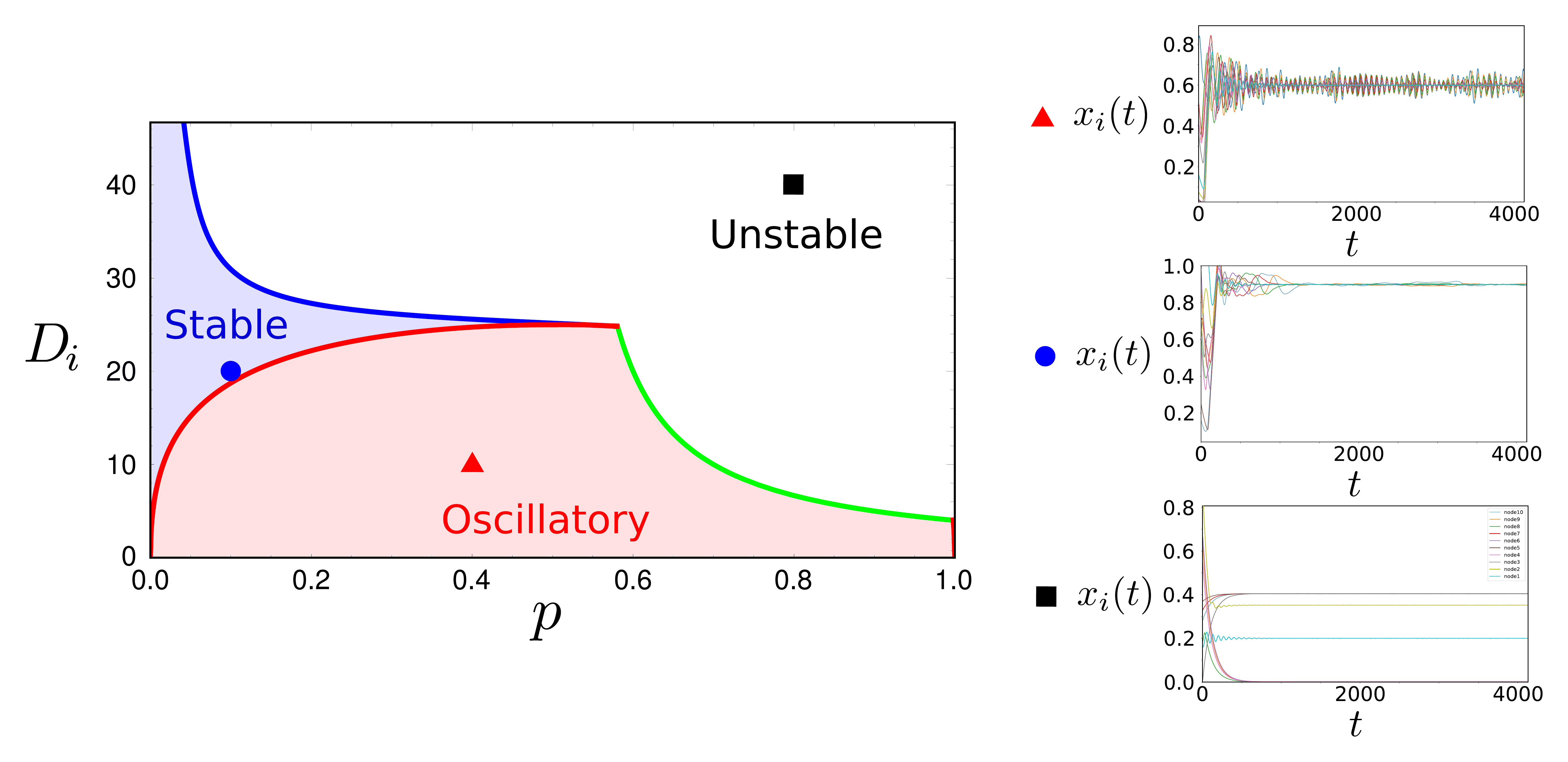}
\caption{\textbf{Left panel}. Stability phase diagram $D-p$, for a fixed value of $r_i=50$ (for all nodes), obtained with the study of eigenvalues of Jacobian \textbf{J}. The continuous lines are the conditions to have stability and oscillations, summarized in the Table \ref{table}. Red line refers condition under which the eigenvalues have imaginary part, indicating presence of oscillations; Blue line refers the condition under which the eigenvalues are real and  negative, indicating a stability of the system. Finally green line refers to the condition under which the eigenvalues are  real and positive. Their intersections create three different spaces of different dynamics.
\textbf{Right panels} Examples of the dynamics of the excitatory activity for a network with $\Omega=10$ nodes and $D_i$ equal for all nodes: $a)$ Oscillations around stable fixed point (red triangle, obtained for $p=0.4$ and $D=10$); $b)$ In the blue region the system is stable without oscillations (blue circle obtained for $p=0.1$ and $D=20$); $c)$ Beyond these two regions The system dynamics is unstable (see black square, obtained for $p=0.8$ and $D=40$).  \label{phase-diagram}}
\end{figure}

The oscillations are related to the imaginary part of the eigenvalues of the $2 \Omega \times 2 \Omega$ Jacobian matrix \textbf{J}, while their real parts are related to the stability of the system \cite{strogatz2018nonlinear}. Here, we look for the region parameters $r_i$, $D_i$, and $p$, leading to oscillations around the stable fixed points. In the Table \ref{table} we summarize the conditions discriminating among different phases of the system (see Appendix A for the analytical derivation), while in Figure \ref{phase-diagram} the phase diagram for this generalized neuromorphic chain is shown. The right panels show examples of the dynamics of the excitatory activity for a network with $\Omega=10$ nodes for the three different possible system's state: stable, unstable and oscillatory. Clearly, the special case $p_i = 1/2$ gives the same result as found in \cite{fanelli2017noise}.
Notice that, when all the parameters have different values for each node, the conditions to have amplified oscillations depend on the specific node index $i$. This means that in order to have amplified oscillation throughout the whole directed chain, these conditions have to be satisfied for all the nodes, otherwise oscillations may arise but then be dumped.

\begin{table}[h]
\begin{center}
\begin{tabular}{|c|c|c|c|c|}
\hline
$p=1/2$ & $p<1/2$ & $p>1/2$ \\
\hline
 & & \\
 &  & if $r_i<r_{min}$ \\
  & & \\
 & & $D_i<\frac{D_0+2r_iz}{\left(1-2p\right)^2}$ \\
  & & \\
\, $D_i<\frac{r_i}{2}$ \, & \, $D_i<\frac{D_0+2r_iz}{\left(1-2p\right)^2}$ \,  & \, if $r_i>r_{min}$ and $p<p_c^1$ or $p>p_c^2$ \,  \\
 & & \\
 & & $D_i<\frac{D_0+2r_iz}{\left(1-2p\right)^2}$ \\
  & & \\
 & & if $r_i>r_{min}$ and $p_c^1<p<p_c^2$ \\
  & & \\
 & & $D_i<\frac{4}{2p-1}$  \\
  & & \\
\hline

\end{tabular}
\caption{Summary table of conditions on $D_i$ and $r_i$, for $p=1/2$, $p<1/2$ and $p>1/2$ to have oscillations around stable fixed point (red region of phase diagrams in Figure \ref{phase-diagram}). In these expressions $z=\sqrt{p(1-p)}$ and $D_0=4p(p-1)r_i$, $p_c^1$ and $p_c^2$ are two critical values of $p$ (intersection points of red and green lines in Figure \ref{phase-diagram}) that are equal for $r_i=r_{min}$. $r_{min}\simeq 13.3$ is the minimum value of $r_i$ in function of $p$, independent of $D_i$. \label{table}}
\end{center}
\end{table}

Since the imaginary part of the eigenvalue of the Jacobian matrix gives the frequency of oscillation of each single node ($\omega_i$), it is possible to find a relation between the distribution of the oscillation frequencies $p(\omega)$ and the one of the non-local coupling constant $p(D)$, constructed from the strengths $D_i$ (for $i=1,2,...,\Omega$) of interactions between units. In fact, by finding an analytical expression of the imaginary part of the system Jacobian eigenvalues $\mathcal{I}m(\lambda_i)=\omega_i=g(D_i)$, it is possible to set the non-local coupling distribution $p(D)$ so to obtain emergent oscillations following a given frequency distribution $p(\omega)$. Formally, the connection is given by $p(\omega)=p(D)\frac{dD}{d\omega}$. 
Focusing on the case $p>p_c^{(1)}>1/2$, $r>r_{min}$, then $D<4/(2p-1)$ and by calculating the eigenvalues of $J$ we obtain  for each node different from the first one (see Appendix A):
\begin{equation}
D(\omega) = \frac{(2 (2 (-1 + p) p r + \sqrt{-4 \omega^2 - (-1 + p) p (r^2 + 16 \omega^2)}))}{(1 - 
  2 p)^2}=g^{-1}(\omega),
\end{equation}
leading to $dD/d\omega = - 8 \omega/(\sqrt{(1 - p) p r^2 - 4 (1 - 2 p)^2 \omega^2})$. In this way, given the distribution of interaction strength $p(D(\omega))$, we can infer the frequency distribution of the system's response to a small perturbation.



\subsection{Adding long-range interactions}

In this section, we add a degree of realism and complexity, by changing the underlying topology of the neuromorphic network. In order to focus on the role of the network topology, we restore the homogeneity constraint, that is  $r_i = r$, $D_i = D$ and $p_i=p=1/2$, $\forall i$. In particular, we want to add long-range interactions among computational units. Long-range interactions are indeed an essential feature in real biological brain \cite{bullmore2012economy},
and from a theoretical point of view is known they facilitate synchronisation of oscillations \cite{barahona2002synchronization}. Onset of synchronisation
triggered by the existence of long-range correlations has been extensively studied for both undirected \cite{barahona2002synchronization} and  directed complex
network topologies \cite{arenas2008synchronization}. Moreover, adding long-range connections in the network breaks
the preferential direction of information transmission of the directed  linear chain. In this way we introduce the presence of feed-back
loops and cycles, that are observed in biological neural circuits and are considered 
essential ingredients to achieve complex computational tasks \cite{dayan2001theoretical}.

The algorithm we implement to introduce long-range connections in the topological structure of the neuromorphic network (given by the matrix $A$) is made of the following steps: \\
$a)$ we fix the directed linear chain (DLC) as a backbone, assigning the weight $A_{i \to i+1}=1$ to each link of DLC;\\
$b)$ with a probability $P$ a new directed link, $a_{i \to j}$, is created, departing from the node $i$ to a node  $j$ which is not connected to $i$ through the DLC. For a fixed $i$ this is done for all $j \in  [ 1, \Omega  ]$, with $j \neq (i,i \pm 1)$.\\
$c)$ this procedure is repeated for each node $i$;\\
$d)$ assign to each $a_{i \to j}$ created in this way the weight $ d \leq 1$.\\

Note that the larger are $P$ and $d$, the higher are the connectivity and the distance from the underlying feedforward acyclic backbone.
In fact, in the way directed weighted neuromorphic networks are built, $d$ measures the relative weights of the random long-range links with respect to the weights associated to the feed-forward $(\Omega-1)$ links of the DLC, while $P$ quantifies their connection density ($P = 1$ corresponds to a fully connected network).

We generate an ensemble of $100$ matrices for each pair $(P, d)$. $P$ and $d$ act as "control parameters": varying their values, different behaviours characterized by given dynamical quantities (amplification, synchronisation and phase locking) can appear. Moreover, for all the analyses here presented we consider $p=0.5$, $D=10$, $r=50$ and $\gamma_i=\gamma=1$. In other words, we select a suitable region where amplified oscillations are observed in the DLC. When $P = 0$ or $d = 0$, we recover the  DLC.

Looking at the stochastic trajectories we identify different situations: 
\begin{enumerate}
\item \textit{Balanced cases} (denoted by the capital letter B):  the system oscillates around the symmetric equilibrium, i.e. the deterministic fixed point at $p=0.5$. We can further split this class in two groups with qualitatively different behaviours:
\begin{itemize}
\item[a.] \textit{Amplification}: amplified oscillations are present, analogous to the one for a DLC. The only difference is that the amplification does not necessarily  increase progressively through the chain, since adding long-range correlations we are changing the natural ranking from $1$ to $\Omega$ of the nodes.
\item[b.] \textit{Synchronisation}: oscillatory nodes share the same dominant frequency, and often  are  phase locked.
 Most importantly, the amplification is sometimes not prevented in this case.
\end{itemize}
\item \textit{Non-Balanced cases} (denoted by NB): the system exhibits non-symmetric fixed points, i.e. different from those obtained for $p=0.5$, and in general different for each node. Again two scenarios can be discriminated:
\begin{itemize}
\item[a.] \textit{Oscillation}: each node oscillates around these non-symmetric fixed points. Moreover, as far as we have observed, the oscillations are almost always synchronous in this situation.
\item[b.] \textit{Convergence}: each node asymptotically converges to these asymmetric fixed points. 
\end{itemize}
\end{enumerate}

\begin{figure}[t]
\includegraphics[width=15 cm]{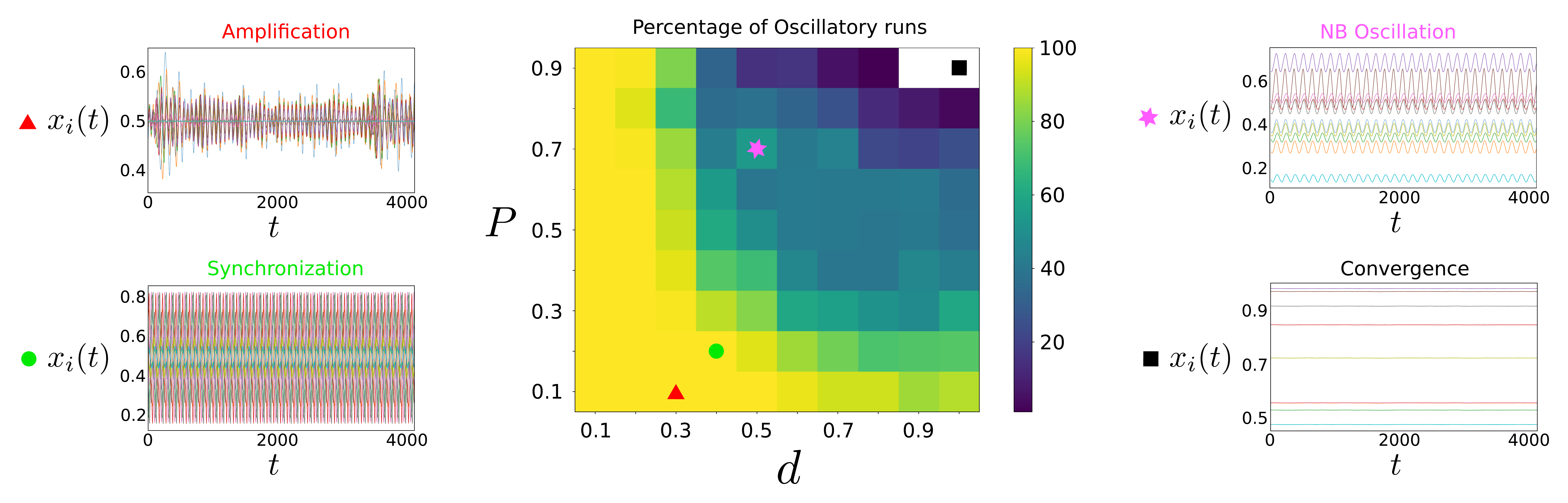}
\caption{
\textbf{Central Panel.} Percentage of oscillatory (both balanced (B) and non-balanced (NB)) cases, as a function of parameters $P$ and $d$ which control the matrix topology.
A run is oscillatory when the mean value of oscillation amplitude of all nodes is higher than a threshold ($10^{-5}$). 
The amplitude of oscillation of each node is measured as the height of principal peak of power spectrum $S_i(\omega)$.
Note that the smaller are $P$ and $d$ (topology close to the DLC), the more frequent are oscillatory cases (yellow squares).
\textbf{Right and left panels.} Some example of different  timeseries $x_i(t)$ are shown, observed for different value of $P$ and $d$, indicated by colored simboles on central phase diagram: (RED TRIANGLE)  a Balanced oscillatory run with amplification; (GREEN CIRCLE) a  Balanced Syncronous run where a collective rhythm emerges; (PURPLE STAR) a Non-Balanced Oscillatory run  and (BLACK SQUARE) a  Non-Balanced Convergent run, observed for high value of $P$ and $d$, when the network has high connectivity.
\label{balance}
}
\end{figure}

In our analysis we investigate the conditions under which emergent amplification and synchronisation of neural activity is observed in neuromorphic networks with long-range interactions. To this aim, we focus on studying only oscillatory cases (both belonging to the group B and NB), since amplification and synchronisation are phenomena intrinsically associated to an oscillatory behaviour, while we do not consider parametrizations that do not give rise to oscillation. In other words, we do not consider convergent NB cases. 

In Figure \ref{balance}, the number of oscillatory runs (for B and NB cases) is shown in the parameter space ($P$, $d$) using a color scale (central panel), along with an example of the dynamics for each behaviour (B cases on the left side and NB cases at right side). Note that the smaller are $P$ and $d$ (i.e. the topology is close to the DLC) and the more frequent are oscillatory cases. The presence of long-range interactions favours the emergence of non-symmetric attractors, around which the nodes can eventually oscillate.

\begin{figure}[t]
\includegraphics[width=17 cm]{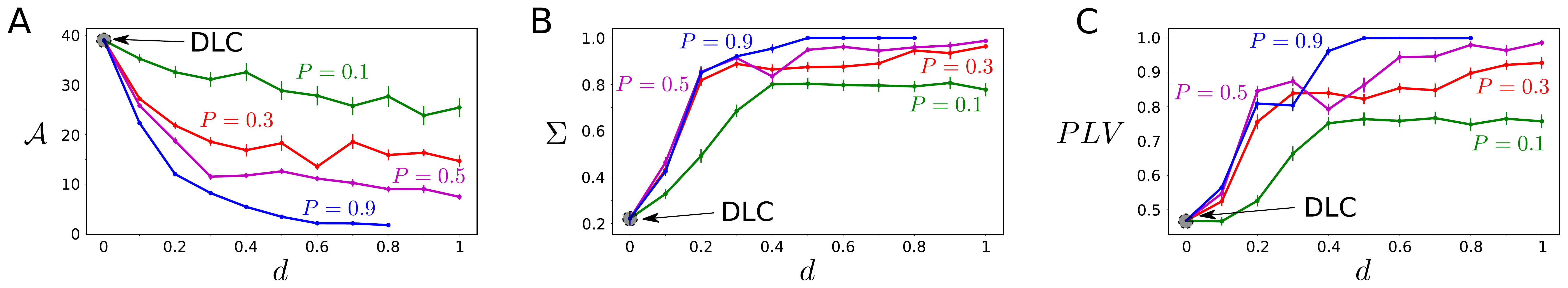}
\caption{
Properties of oscillatory  cases, as a function of strength of long-range connections $d$, at different value of probability $P$ to add long-range. 
A) Amplification factor $\mathcal{A}$ between nodes, evaluated as in  equation (\ref{ampl}), averaged over oscillatory cases of ensamble matrices generated for each couple of $P$ and $d$. We observe that the amplification factor becomes less strong as we increases the parameters $(P,d)$. For each point also the standard error is shown. Note that, for comparison the  amplification factor averaged over 100 runs for the DLC is shown for $d=0$ and it is about 40 $ dB$ .
B) Global synchronisation in frequency $\Sigma$ evaluated as in equation (\ref{sync}), averaged over oscillatory cases. The probability to have a  synchronous run increases with $(P,d)$.
C) Phase Locking Value calculated as the mean of all PLV of each pair of nodes as defined in equation (\ref{PLV}), averaged over oscillatory cases. In agreement with the trend of synchronisation in frequency, the probability to have phase coherence increases with $P$ and $d$. This emergent phenomenon is absent in the DLC.  
\label{transition}
   }
\end{figure}

In Figure \ref{transition} we show some dynamical quantities of interest as a function of the network parameters $d$ and $P$. Each point corresponds to the average over the number of oscillatory cases in the matrix ensemble generated for the corresponding pair $(P,d)$. 

For each time series, we estimate the amplification of the signal in decibel as follows:

\begin{equation}
\mathcal{A}=10 \log_{10} \frac{\max_i(\sigma_i)}{\min_i(\sigma_i)}
\label{ampl}
\end{equation}  
where $\sigma_i$ is the dominant peak height of the power spectrum associated to the node $i$. We note that a similar quantification can be made using the  variance of the time-series of each run, exhibiting the same result (see Appendix B).
We find that the amplification factor $\mathcal{A}$ decreases on average with $P$ and $d$ (Panel A), meaning that all the nodes tend to oscillate with the same amplitude
as we depart from the linear chain.

Looking at the single run  stochastic time series and its power spectum  (see Appendix C), we note that in the DLC not all the oscillatory nodes share the same dominant frequencies,  while increasing $P$ and $d$ we observe an increase of frequency entrainment.
  A cluster of synchronous nodes oscillating with the same collective frequency appears. It can involve all the $\Omega$  nodes (as
  the run shown in Figure \ref{S3}B in Appendix C) or a subset of them (excluding only those showing negligible amplitude of oscillation).   
We therefore introduce a continuous function quantifying the network frequency synchronisation as follows:
\begin{equation}
\Sigma=\frac{1}{\Omega} \sum_{i=1}^{\Omega} \frac{1}{1+(\frac{\Delta_i}{\Delta_0} )^2}
\label{sync}
\end{equation}
where $\Delta_i=\omega_i - \omega_{mean}$ measures the distance of the dominant frequency of node $i$ ($\omega_i$) from the mean frequency of all nodes ($\omega_{mean}$)
weighted over nodes oscillations amplitudes, i.e., $\omega_{mean}= \frac{\sum_{i=1}^{\Omega} \omega_i \sigma_i}{\sum_{i=1}^{\Omega} \sigma_i}$.
Clearly for perfect collapse of frequencies $\Sigma=1$, while $\Delta_0$ sets the degree of frequencies dispersion required to get $\Sigma=1/2$.
For DLC, equation (\ref{sync}) gives a low global synchronisation  $\Sigma=0.2$,
in agreement with the observation that the  dominant peaks do not share the same frequency.
The global frequency synchronisation value averaged over all oscillatory runs (Panel B of Figure  \ref{transition}) increases with $(P,d)$.

From the opposite trends of average $\mathcal{A}$ and $\Sigma$, we conclude that, increasing $(P,d)$, the preferred directionality of the information spreading through the network is gradually lost in favour of a uniform diffusive response.
In other words,  once the long-range interactions reach a given concentrations, they tend to dominate over the topological backbone constituted by the DLC.

Inspired by the crucial role of noise-robust phase locking  in real neural network \cite{varela2001brainweb,siegel2009phase}, we further consider the degree of phase-locking synchronisation or phase coherence. We then measure the phase locking value (PLV) of  a pair of network nodes $j$ and $k$,  as \cite{boccaletti2018synchronization}
\begin{equation}
PLV_{jk}= <|e^{I (\phi_j(t)-\phi_k(t))}|> 
\label{PLV}
\end{equation}
  where $I$ is the imaginary unit,  $< >$ the average over  time, and $\phi_j(t)$ the istantaneous phase variable of node $j$  at time t, defined as $\phi_j(t)=arctan (\frac {\bar x_j(t)}{x_j(t)}) $ where ${\bar x_j(t)}$ is the Hilbert transform of the observed time series $x_j(t)$ of node $j$.
  Clearly a uniform distribution of phase difference $\phi_j(t)-\phi_k(t)$ will results in  $PLV_{jk}=0 $, while $PLV_{jk}$ reaches its maximum value
 $PLV_{jk}=1$ if nodes are strictly phase locked ($\phi_j(t)-\phi_k(t)$ is constant in time).

  In Panel C of Figure \ref{transition} we show the average of $PLV_{jk}$ over both node indices, named $PLV$, again averaged over all oscillatory cases for each point ($P,d$).
  
We find a very strong correlation between $\Sigma$ and PLV, meaning that neuromorphic networks with long-range interactions whose dynamics has strong frequency entrainment, often exhibit a  strong phase coherence.

\section{Discussion and Conclusion}

We have shown how adding long-range interactions in a neuromorphic directed network leads to novel emergent phenomena in the system dynamics that is not observed in the chain, namely synchronisation and phase coherence of the artificial neurons activity. 

Figure \ref{transition} notably shows how there exists a region of the network parameters where amplification and synchronisation (both in the dominant frequency and in the phase coherence) coexist. These neuromorphic networks are characterized by small values of $P$ and large enough $d$, meaning that few but strong long-range connections can trigger a global synchronisation, while preserving the amplification of the signal.
On the other hand we have found that the mean amplification factor (as measured by $\mathcal{A}$) decreases when $d$ increases, for any values of $P$ (Panel A of  Figure \ref{transition}).  The observed decrease of the amplification factor (for oscillatory runs) reflects an increase in the concerted activity of the units even among distant nodes as the network starts to be more and more connected.
This collective behaviour is also reflected in the increase of the probability that all nodes share a  global collective dominant frequency and phase. In fact, both $\Sigma$ and $PLV$ increase for increasing $d$ (and $P$) as shown in Panels B-C of  Figure \ref{transition}.
A relevant important feature emerging from the statistics is therefore that most of the runs become synchronous as the long-range interactions get more frequent and more strong. This again evidences the prominent role of long-range connections in triggering synchronisation.

\begin{figure}[h]
\begin{center}
\includegraphics[width=10 cm]{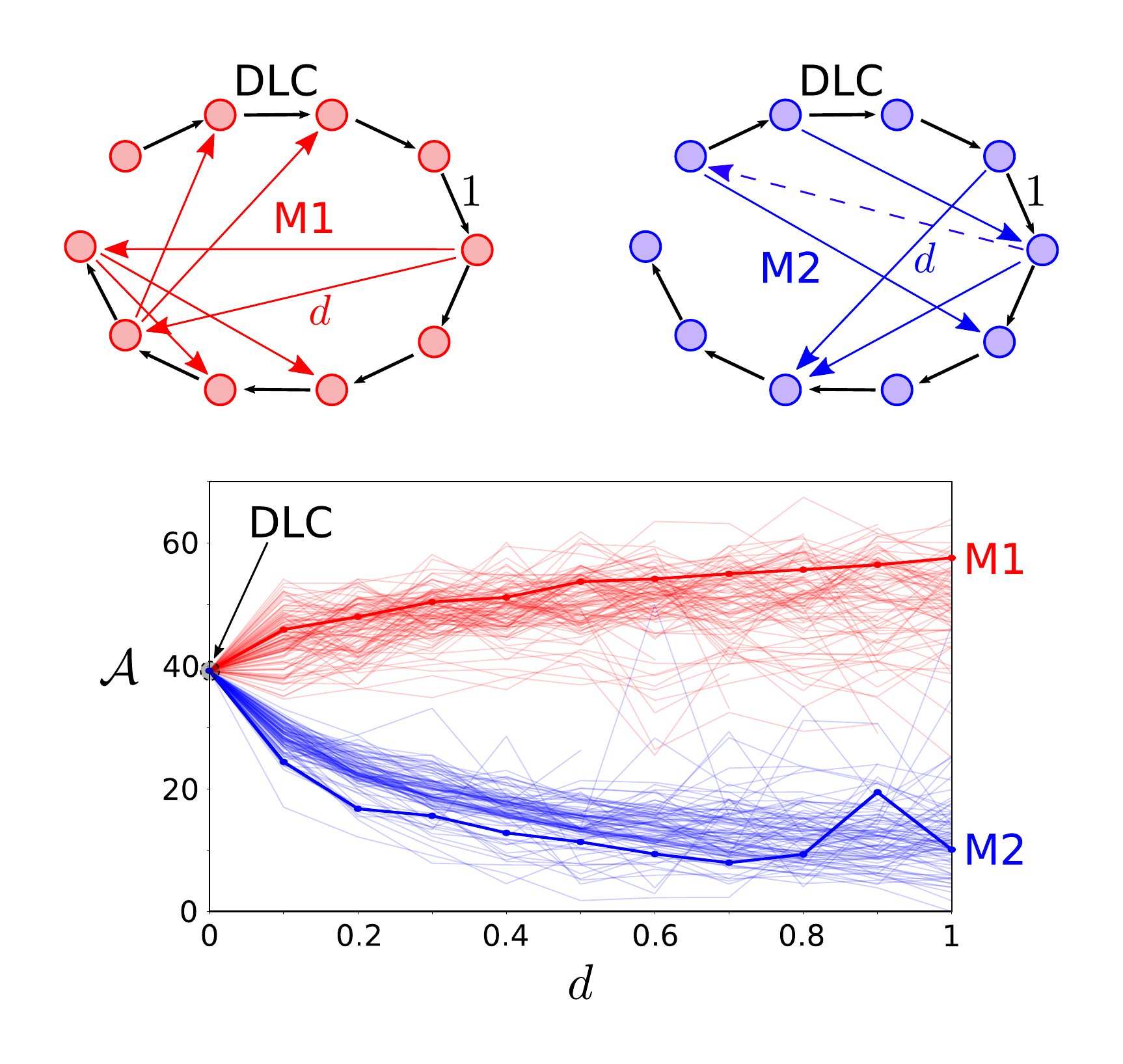}
\caption{
TOP.  A schematic representation of two networks, M1 (red) and M2 (blue), created with $P=0.1$ and $d=0.5$ with no ingoing link in the first node (M1) and with a link in input in the first node from another one of the network (M2). 
BOTTOM. Amplification factor in function of strength of connections $d$ for the single network M1 (red line) and for the single network M2 (blue line). This result shows that the super local topological feature of network M1 (no link ingoing in  the first node) gives an amplification factor higher than the directed linear chain and increasing with $d$. On the contrary, the amplification factor of network M2 decreases with $d$, following the same trend of amplification factor of ensemble matrices in Figure \ref{transition}. This means that the networks like the network M1 are rare when we create an ensemble of matrices for each pair of $(P,d)$. Moreover, if we create two ensemble of 100 matrices with the same $P=0.1$ for each value of $d$ with the super local feature of first node like in the network M1,  we obtain the same trend of single matrix M1 in function of $d$ (shown in opaque, under the red line of single matrix M1), confirming the effect of this topological feature on the amplification factor. The same happens with an ensemble networks like network M2, that follows the same specific trend of single matrix M2.
\label{good}
}
\end{center}
\end{figure}

One may be tempted to think that the main reason of observed decreased in amplification while departing from the linear chain, is because by adding long-range interactions with backward direction is decreasing the non-normality of the network \cite{asllani2018structure}. In fact, as suggested by \cite{fanelli2017noise}, one possible explanation is that the asymmetry in the imposed couplings yields the observed non trivial amplification mechanism in the linear chain. Therefore, as far as the network of couplings results in a non-normal adjacency matrix, then giant cycles across the networks can be observed. However, by adding backward long-range interactions, we are making the network more symmetric and we are breaking the non-normality thus causing the decrease in the amplification.

However, we found that this is not the case. In fact, among the generated neuromorphic networks, we found some "special" matrices for which the amplification factor improves with respect to the DLC for increasing $d$! A comparison between two different matrices $M1$ and $M2$ (chosen among those generated with long-range connection with probability $P=0.1$) is presented in Figure \ref{good}. The amplification factor has two completely different behaviours as a function of $d$, as shown in the lower diagram of Figure \ref{good}.

Investigating on the reason of this particular result, we found that the crucial ingredient discriminating between the two cases is a very local feature of the network topology. In fact we found that if the neuromorphic network has the in-degree of the first node of the backbone equal to zero, i.e., the first node does not have incoming links, then its amplification increases with $d$ and is higher with respect to the DLC. 

To confirm this result, we generate an ensemble of 100 matrices for $P=0.1$ and different $d$, imposing in one case (red lines) a zero in-degree of the first node of the backbone, $k^1_{in}=0$, while in the other (blue lines) $k^1_{in}>0$. The related amplification factors $\mathcal{A}$ are shown in the main panel of Figure \ref{good}: in the first cases $\mathcal{A}$ increases with $d$ and it is on average larger than the amplification of the DLC, while in the second cases the opposite is true, i.e. $\mathcal{A}$ decreases with $d$ and it is on average smaller than the one of the DLC.
Clearly, when we generate the neuromorphic network by adding at random places  long-range interactions (as in Figure \ref{transition}), having the local topological feature $k^1_{in}=0$ becomes more and more rare as $P$ increases.

It is important to observe that for these "special" networks with $k^1_{in}=0$ both the synchronisation and phase locking value do not change their trend as a function of $d$ (see Appendix B). In these way we can generate neuromorphic network with higher amplification of the DLC, but also where synchronisation and phase coherence emerge.
Here we want to stress that the simultaneous appearance of a synchronous and amplified signal can ideally lead to a robust and coherent transmission of information, which is believed to be a feature of paramount importance in real neural networks \cite{varela2001brainweb,siegel2009phase,uhlhaas2009neural,euston2007fast}.
Therefore, with this work we propose  the core topological ingredients needed to design neuromorphic networks which can perform complex computational tasks with simple dynamical rules.

\section{Acknowledgements}
S.Suweis ack STARS ReACT UNIPD grant.

\newpage

\section*{Appendix A: Stability conditions of generalized directed linear chain}

Here we derive the conditions reported in Table \ref{table} of the main text, characterizing the stability phase diagram of the DLC with heterogeneity in the model parameters. To this aim, we study the spectrum of the eigenvalues of Jacobian matrix \textbf{J} of equation (\ref{zeta}) in the main text.

In particular, we consider a DLC costituted by $\Omega$ nodes and the dynamical local and non-local parameters $r_i$ and $D_i$ dependent on the single node $i$, while the parameter $p_i=p$ equal for all nodes. Moreover we remember that $r_i$ and $D_i$ are positive quantities, and $0<p<1$.  

In these conditions the arguments of sigmoid function (equations (\ref{s})) are:
\begin{eqnarray}
s_{x_i}&=&-r_i\left(y_i-p\right)+D_i\left(x_{i-1}-x_i\right)-D_i\left(y_{i-1}-y_i\right) \nonumber \\
s_{y_i}&=&r_i\left(x_i-(1-p)\right)+D_i\left(x_{i-1}-x_i\right)-D_i\left(y_{i-1}-y_i\right)
\label{s1}
\end{eqnarray}

Thanks to the ansatz of Linear Noise Approximation (LNA) \cite{van1992stochastic}, the noise acts as a first order perturbation to the deterministic fixed points, that is $x_i=x^{*}+\frac{\xi_i}{\sqrt{V_i}}$ and $y_i=y^{*}+\frac{\eta_i}{\sqrt{V_i}}$. Performing the expansion up to the first order in $\frac{1}{\sqrt{V_1}}$ in the equations (\ref{langevin}), we end up with the Langevin equation (\ref{zeta}) for the fluctuations $\vec{\zeta} = ( \vec{\xi}, \vec{\eta} )$
where $\textbf{J}$ is:

\begin{center}
\begin{equation}
J =
	\begin{bmatrix}
	E_1 & 0 & 0 &  ... & 0 \\
	S_2 & E_2 & 0 & ... & 0 \\
	0 & S_3 & E_3 & ... & 0 \\
	\vdots & \vdots & \ddots & \ddots & \vdots \\
	0 & 0 & ... & S_{\Omega} & E_{\Omega} 
	\end{bmatrix}
	\label{J}
\end{equation}
\end{center}

with 

\begin{center}
\begin{equation}
	E_1 = 
	\begin{bmatrix}
	-1 & -\frac{r_i(1-p)}{2} \\
	\frac{r_i p}{2} & -1
	\end{bmatrix}
\end{equation}
\end{center}

\begin{center}
\begin{equation}
	E_i	= \frac{1}{\gamma_i}
	\begin{bmatrix}
	-1-\frac{D_i(1-p)}{2} & -\frac{(r_i-D_i)(1-p)}{2} \\
	\frac{(r_i-D_i)p}{2} & -1+\frac{D_ip}{2}
	\end{bmatrix}
\end{equation}
\end{center}

and for  $i\geq2$

\begin{center}
\begin{equation}
	S_i	= \frac{D_i}{2\sqrt{\gamma_i\gamma_{i-1}}}
	\begin{bmatrix}
	1-p & -(1-p) \\
	p & -p
	\end{bmatrix}
\end{equation}
\end{center}

We remember that in order to have oscillations around stable fixed point the Jacobian matrix \textbf{J} must have complex eigenvalues with negative real part. If it has negative real eigenvalues, the system converges to a stable fixed point. In this case, the eigenvalues of \textbf{J}, equation \eqref{J}, corresponde to the eigenvalues of matrices $E_i$ on its diagonal.

The eigenvalues of matrix $E_1$ are:
\begin{equation}
\lambda_{1,2}=-1 \pm\ I\frac{r_i}{2}\sqrt{p(1-p)}
\end{equation}
always complex $\forall$ $r_i$ ($I$ is the imaginary unit) and with negative real part, allowing for oscillations around the stable fixed point for the node $1$. The eigenvalues of matrices $E_i$ (for $i\geq2$) are: 
\begin{equation}
\lambda_{3,4}^i=\frac{1}{\gamma_i}\left[-1+\frac{2p-1}{4}D_i\ \pm \frac{1}{4}\sqrt{\left(1-2p\right)^2D_i^2-8\left(p-1\right)pD_ir_i+4\left(p-1\right)pr_i^2}\right]
\label{eigenvalues}
\end{equation}

The system is stable when the eigenvalues are real and negative. In our case, $\lambda_{3,4}^i<0$ (negative, but not necessarily real) when the following condition holds:

\begin{equation}
D_i<\frac{-4-pr_i^2+p^2r_i^2}{2(1-2p-pr_i+p^2r_i)}
\label{stable}
\end{equation}

The blue line in Figure \ref{phase-diagram} identifies the set of points for which this latter conditions is satisfied as an equality.

If the argument in square root of eigenvalues  (\ref{eigenvalues}) is negative we can express the eigenvalue as a complex number $\lambda_{3,4}^i=\mathcal{R}e\left\{\lambda\right\}+I \mathcal{I}m\left\{\lambda\right\}$, where

\begin{equation}
\mathcal{R}e\left\{\lambda\right\}=\frac{1}{\gamma_i}\left(-1+\frac{2p-1}{4}D_i\right)
\end{equation}

and

\begin{equation}
\mathcal{I}m\left\{\lambda\right\}=\frac{1}{4\gamma_i}\sqrt{-\left(1-2p\right)^2D_i^2-8\left(1-p\right)pD_ir_i+4\left(1-p\right)pr_i^2}
\label{freq_risonanza}
\end{equation}

which gives the frequency of oscillations of each single node $i$,  $\omega_i$, as a function of parameters $p$, $r_i$ and $D_i$. The condition on $D_i$ to have a negative argument in the square root of $\lambda_{3,4}^i$, triggering oscillations, is:

\begin{equation}
\frac{D_0-2r_iz}{\left(1-2p\right)^2}<D_i<\frac{D_0+2r_iz}{\left(1-2p\right)^2}
\label{condDi}
\end{equation}

where $z=\sqrt{p(1-p)}$ and $D_0=4p(p-1)r_i$. $D_0$ is always negative because $p<1$; choosing $D_i>0$, the condition (\ref{condDi}) becomes:

\begin{equation}
0<D_i<\frac{D_0+2r_iz}{\left(1-2p\right)^2}
\label{condDi2}
\end{equation}

The condition (\ref{condDi2}) is the condition on $D_i$ to have a complex eigenvalue. In what follows we derive the conditions to have also a negative real part of (\ref{eigenvalues}), thus identifying the regions in which the system oscillates around its stable fixed point.

We observe that for $p<1/2$ the real part of $\lambda_{3,4}^i$ is always negative, so the only condition to have oscillations around stable fixed point is the (\ref{condDi2}), whose upper bound is represented in Figure \ref{phase-diagram} by a red line. Clearly this bound is tighter than the one in equation \eqref{stable}.

On the contrary, for $p>1/2$ the condition to have negative real part is:

\begin{equation}
D_i<\frac{4}{2p-1}
\label{condDi3}
\end{equation}

The conditions (\ref{condDi2}) and (\ref{condDi3}) are the two conditions to have oscillations around stable fixed point for $p>1/2$, that we can consider as two function of $p$ and dependent on the choice of $r_i$. Merging the two, the $r_i$ lying on both upper bounds changes as a function of $p$ as follows:
\begin{equation}
r_i=-\frac{2\left(2-2p+\sqrt{-1+\frac{1}{p}}\right)}{1-3p+2p^2}
\label{ri}
\end{equation}

This equation has two solutions, $p_c^1$ and $p_c^2$, and a minimum $r_{min}$. This means that the condition in equation \eqref{condDi3} effectively applies only when $p_c^1 < p < p_c^2$ and $r > r_{min}$. Its upper bound is represented by the green line in Figure \ref{phase-diagram}.

To summarize the discussion above for $p>\frac{1}{2}$, we can conclude that: 
\begin{enumerate}
\item for $r>r_{min}$ 
\begin{itemize}
\item for $p<p_c^1$ and $p>p_c^2$ 
\begin{equation}
D_i<\frac{D_0+2r_iz}{\left(1-2p\right)^2}
\end{equation}

\item for $p_c^1<p<p_c^2$ 
\begin{equation}
D_i<\frac{4}{2p-1}
\end{equation}
\end{itemize}

\item for $r \leq r_{min}$ 
\begin{equation}
D_i<\frac{D_0+2r_iz}{\left(1-2p\right)^2}
\end{equation}
\end{enumerate}

Finally, for $p=1/2$ the eigenvalue $\lambda_{3,4}^{i}$ becames:

\begin{equation}
\lambda_{3,4}^{i}=\frac{1}{\gamma_i}[-1 \pm \frac{1}{4}\sqrt{2D_ir_i-r_i^2}]
\end{equation}
We can write $\lambda_{3,4}^i=\mathcal{R}e\left\{\lambda\right\}+ I \mathcal{I}m \left\{\lambda\right\}$ if $2D_ir_i-r_i^2<0$, i.e. 

\begin{equation}
D_i<\frac{r_i}{2}
\label{pmezzo}
\end{equation}

The real part $\mathcal{R}e\left\{\lambda\right\}=-1$ is always negative.

The results of this section are summarized in Table \ref{table} in the main text.

\section*{Appendix B: Sensibility analysis}

  To  check the robustness of the results, the amplification factor, evaluated as in (\ref{ampl}) using heights of dominant peaks, is compared with a measure based on the variance of time series of each node, $v_i$, namely:
%

\begin{equation}
\mathcal{\bar A}= 10 \log_{10} \left( \frac{\max_i(v_i)}{\min_i(v_i)} \right)
\label{ampl2} 
\end{equation}

\begin{figure}[h]
\begin{center}
\includegraphics[width=15 cm]{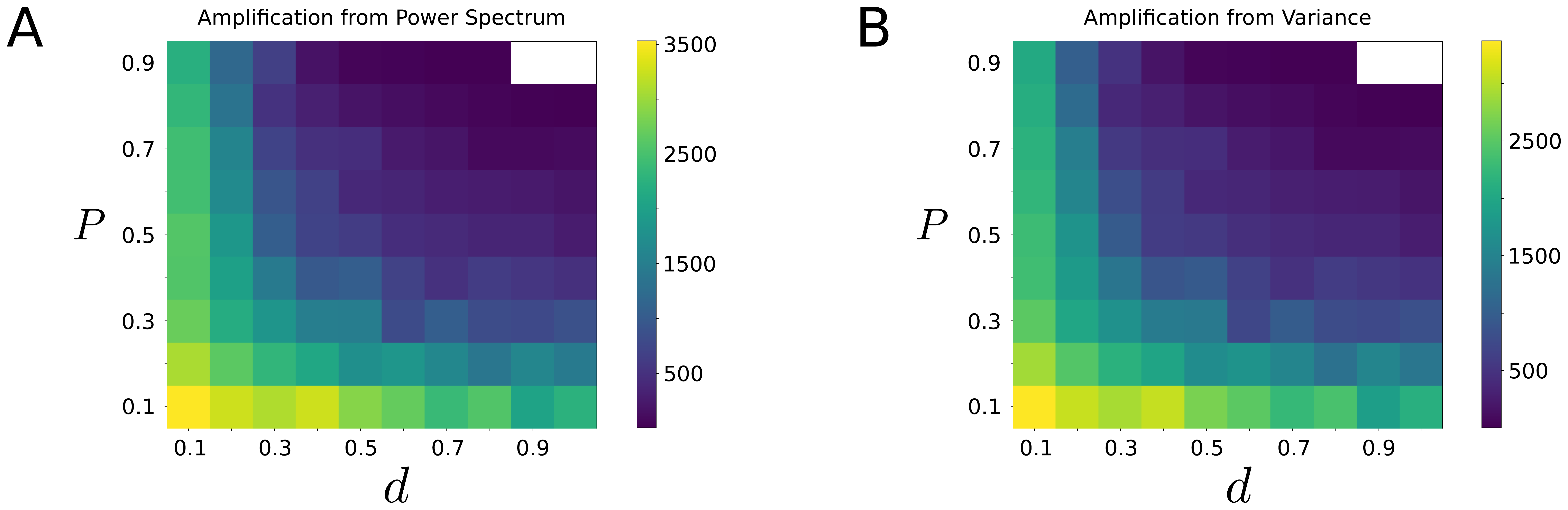}
\caption{ Comparison of Amplification factor using two different measures, shown in function of $P$ and $d$ for all oscillatory runs.
  The amplitude of oscillations is evaluated in two different ways: in Panel A it uses the height of dominant peak of power spectrum, as in equation (\ref{ampl}); in Panel B it uses the variance of time series as in equation (\ref{ampl2}). The two measures gives qualitatively equivalent result.}
\label{S2}
\end{center}
\end{figure}

The results in Figure \ref{S2} are evaluated averaging over all oscillatory runs, and  show that both measures (equation \eqref{ampl} and equation \eqref{ampl2}) give consistent behaviour.

\begin{figure}[h]
\begin{center}
\includegraphics[width=16 cm]{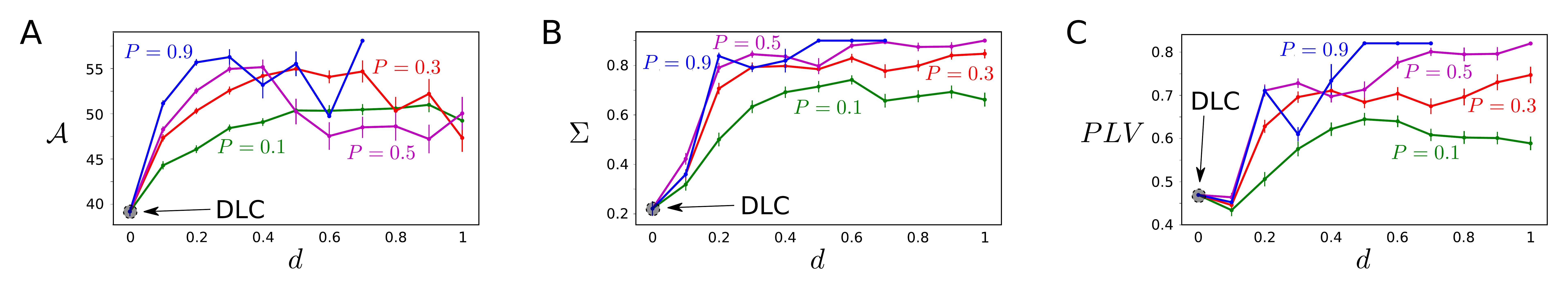}
\caption{ Amplification (Panel A),   frequency  synchronisation (Panel B) and Phase Locking Value (Panel C) in function of $d$ and for different values of $P$, averaged over oscillatory cases of an ensamble of matrices with a zero in-degree,  $k^1_{in}=0$, of the first node of the backbone, as M1 in Figure \ref{good}.  In agreement with the case $P=0.1$, shown in Figure \ref{good}, the amplification factor for such "special" matrices is higher than for the DLC, and it is completely different from the trend shown in Figure \ref{transition}A. However the global synchronisation (Panel B) and PLV (Panel C), averaged over oscillatory runs of such special matrices, do not change their trend as function of $d$  with respect to that obtained in Figure \ref{transition}B-C. The values of these three dynamical quantities for the DLC is also drawn, in order to show how this "special" ensamble of matrices  generate neuromorphic network with higher amplication of the DLC, but also where synchronisation and phase coherence (absent in DLC) emerge.}
\label{S4}
\end{center}
\end{figure}

We then focus on the dynamical features observed for matrices with zero in-degree of the first node of the backbone, as matrix M1 in Figure \ref{good}.
The amplification factor, the frequency synchronisation and phase locking value are shown in Figure \ref{S4}. They are evaluated averaging over all oscillatory
runs of an ensamble of 100 connectivity matrices for each pair $(P,d)$ built with the constraint of zero in-degree of first node $k^1_{in}=0$.
As expected looking at the opaque red lines in Figure \ref{good}, the amplification factor shown in Figure \ref{S4} for the “special” matrices increases with respect to the DLC. Figure \ref{S4} underlines that the increase of amplification shown in Figure \ref{good} is robust whit respect to different values of $P$ and $d$.

  Notably the frequency synchronisation and the phase locking coherence also increase with respect to  DLC  with $(P,d)$,
with a trend in agreement with that observed  for the set of all matrices without the condition $k^1_{in}=0$, shown in Figure \ref{transition}. 

This  confirms that the analysis is not fine tuned and underlines the robustness of our results. 

\section*{Appendix C: Single run stochastic Time Series, Power Spectrum  and Synchronization}

\begin{figure}[t]
\begin{center}
\includegraphics[width=16 cm]{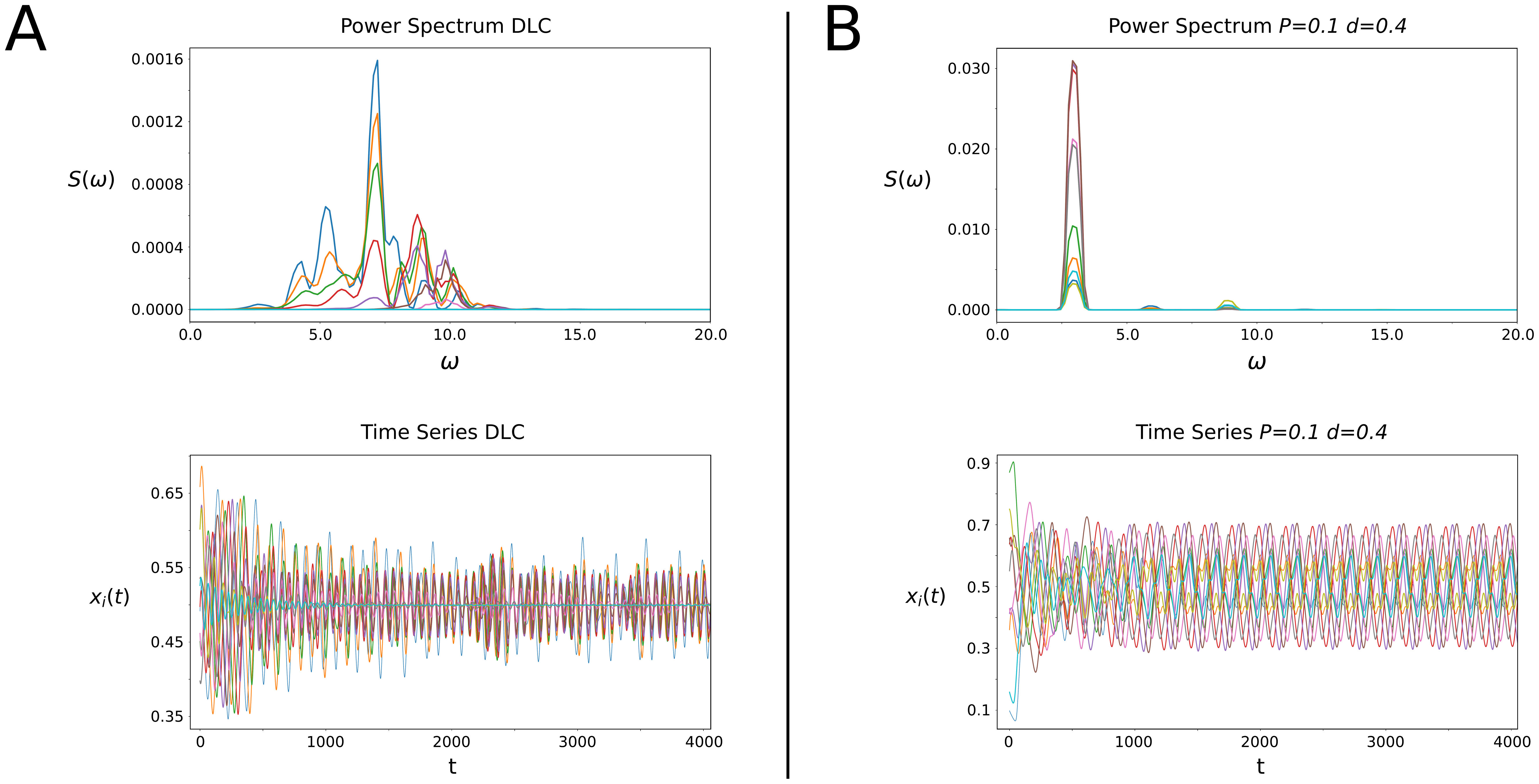}
\caption{Power Spectrum and Time Series of excitatory nodes of Directed Linear Chain (\textbf{Panel A}) is compared with the dynamics of the  directed weighted neuromorphic network with $P=0.1$ and $d=0.4$ (\textbf{Panel B}). The power spectrum of the DLC shows how the nodes do not have the same dominant frequency. Moreover from the time series we found that there is not phase locking synchronisation. On the contrary, in the run with long-range connections  $P=0.1$ and $d=0.4$ on the right, we found that all nodes share the same dominant frequency, and the time series shows that, after a short transient, a collective phase-locked rhythm emerges, a novel feature not observed in the DLC.}
\label{S3}
\end{center}
\end{figure}

Looking at single runs of the stochastic trajectories of nodes activity ($x_i(t),y_i(t)$ for $i=1,...,\Omega$) we observe a qualitative difference between the DLC and the neuromorphic network with long-range connections.
In Figure \ref{S3}A we show the time series of excitatory nodes in a  single run of the DLC dynamics.
In the DLC, the first node (and some of nearby nodes) has a  dominant frequency different from the others, while a strong frequency entrainment and phase coerence can appear in the  neuromorphic network, as shown for the run with $P=0.1, d=0.4$ 
in Figure \ref{S3}B.
In the neuromorphic networks  with long-range links, we often observe the emergence of a cluster of synchronous and phase coeherent oscillations, that can involve all
the $\Omega$ nodes as in  Figure \ref{S3}B or only a subset of nodes (while the other nodes do not oscillate).
Note that the emergence of synchronisation and phase coeherence do not prevent to have a good amplification among nodes, as seen from the gradual increase of amplitude of
oscillation of nodes in the $P=0.1, d=0.4$ case shown in  Figure \ref{S3}B.
For the runs shown in  Figure \ref{S3}B, the global frequency synchronization defined in equation (\ref{sync}) is $\Sigma= 1.0$, while the phase coherence  defined
in equation (\ref{PLV}) is PLV$=0.95$.
The quantities $\Sigma$ and PLV (averaged over an ensable of 100 matrices) are shown as a function of $P$ and $d$ in the main text, Figure  \ref{transition}.

\end{document}